\documentclass[prl,aps,twocolumn,superscriptaddress]{revtex4-1}
\usepackage{graphicx,color}
\usepackage{amsthm}
\usepackage{amsfonts}
\usepackage{algorithmic}
\usepackage{enumerate}
\usepackage{latexsym}
\usepackage{amsmath}
\usepackage{amssymb}
\usepackage[colorlinks=true,citecolor=blue,linkcolor=blue]{hyperref}

\newcommand{\sigmadc}{\sigma_{\rm dc}}
\emergencystretch=\maxdimen
\newcommand{\be}{\begin{equation}}
\newcommand{\ee}{\end{equation}}
\newcommand{\ba}{\begin{eqnarray}}
\newcommand{\ea}{\end{eqnarray}}

\newcommand{\COMMENTED}[1]{}

\newcommand{\ob}[1]{{\langle #1\rangle}}

\newcommand{\bfi}{\mathbf{i}}
\newcommand{\bfj}{\mathbf{j}}

\newcommand{\bfr}{\mathbf{r}}

\newcommand{\bfQ}{\mathbf{Q}}

\newcommand{\bfS}{\mathbf{S}}

\newcommand{\hH}{{\hat{H}}}

\newcommand{\hc}{{\hat{c}}}

\newcommand{\hj}{{\hat{j}}}
\newcommand{\hn}{{\hat{n}}}

\begin{document}

%% \title{Disorder-driven correlated Anderson insulator 
%% in the Hubbard model on honeycomb lattices}
\title{A correlated Anderson insulator on the honeycomb lattice}

\author{Tianxing Ma}
\affiliation{Department of Physics, Beijing Normal University, Beijing
100875, China\\}
\affiliation{Beijing Computational Science Research Center, Beijing
100193, China}
\author{Lufeng Zhang}
\affiliation{Department of Physics, Beijing Normal University, Beijing
100875, China\\}
\author{Chia-Chen Chang}
\affiliation{Department of Physics, University of California, Davis,
California 95616, USA}
\author{Hsiang-Hsuan Hung}
\affiliation{Department of Physics, Beijing Normal University, Beijing
100875, China\\}
\affiliation{Department of Physics, The University of Texas at Austin,
Austin, TX, 78712, USA }
\author{Richard T. Scalettar}
\affiliation{Department of Physics, University of California, Davis,
California 95616, USA}

\begin{abstract}
We study the effect of disorder on the semimetal -- Mott insulator
transition in the half-filled repulsive Hubbard model on a honeycomb
lattice, a system that features vanishing density of states at the 
Fermi level. Using the determinant quantum Monte Carlo method, we 
characterize various phases in terms of the bulk-limit antiferromagnetic 
(AF) order parameter, compressibility, and temperature-dependent DC 
conductivity. In the clean limit, our data are consistent with previous 
results showing a single quantum critical point separating the semi-metallic 
and AF Mott insulating phases. With the presence of randomness, a non-magnetic 
disordered insulating phase emerges. Inside this disordered insulator phase, 
there is a crossover from a gapless Anderson-like insulator to a gapped 
Mott-like insulator.
\end{abstract}

\pacs{71.10.Fd, 74.20.Rp, 74.70.Xa, 75.40.Mg}

\maketitle

% --------------------------------------------------------------------
\noindent
\underline{\it Introduction} ---
The study of the metal-insulator transition (MIT) has a long history.  A
classification of metals and insulators based on band theory was
established in the early years of quantum
mechanics.\cite{Bethe1928,*Sommerfeld1928,*Bloch1929} After the
discovery of transition-metal oxides\cite{deBoer1937} (e.g., NiO) where
the $d$-orbitals are partially filled, it was realized that the band
theory is insufficient and interactions between electrons should be
taken into account.\cite{Mott1937} The resulting Mott insulator has 
become a paradigm for the physics of MIT in strongly correlated
systems.\cite{Imada1998}
While non-interacting systems typically show metallic properties,
Anderson, in a seminal work, showed that, in the presence of strong
disorder, electron eigenstates can be localized and fall off exponentially
with distance due to coherent
backscattering.\cite{Anderson1958,*Anderson1978} This phenomenon has
been verified by experiments,\cite{Wiersma1997,Storzer2006} and the
Anderson localization mechanism provides a third route to the
metal-insulator transition.

In real materials, since disorder and interactions are both present,
understanding the interplay of the two sources of localization and their
combined impact on the MIT has become a focus of 
research\cite{Curro2009,Keimer2015,Dagotto2005,Kravchenko1994}. 
On the theory side, progress has been made by treating correlations at the
Hartree-Fock level\cite{Lee1985,Belitz1994}
and via diagrammatic\cite{Lee1985} and perturbative renormalization
group calculations\cite{Finkelstein1983,Castellani1984,Belitz1994}. However, 
questions remain when both disorder and interactions are strong,\cite{Imada1998} 
in which case they should be treated on the same non-perturbative footing.
New theoretical concepts challenging existing paradigms have
also emerged.\cite{Nandkishore2015} For experiments, recent progress in
controlling disorder and interaction parameters precisely is allowing
detailed comparison with theoretical predictions.\cite{Pasienski10,*Kondov2015,*Schreiber2015,*Bordia2015}

In order to shed light on the physics of disordered interacting fermions, we 
study the half-filled repulsive Hubbard model on the honeycomb lattice with 
off-diagonal disorder using the numerically exact determinant quantum Monte 
Carlo (DQMC) method.\cite{WhHbk}  While most previous DQMC and other numerical 
studies have examined the interplay of disorder and correlations in `conventional' 
geometries with a finite density of states at the Fermi level $E_F$, our study 
addresses the interesting issue of how this interplay manifests within a 
Dirac-like dispersion near $E_F$.

\begin{figure}
\includegraphics[scale=0.4]{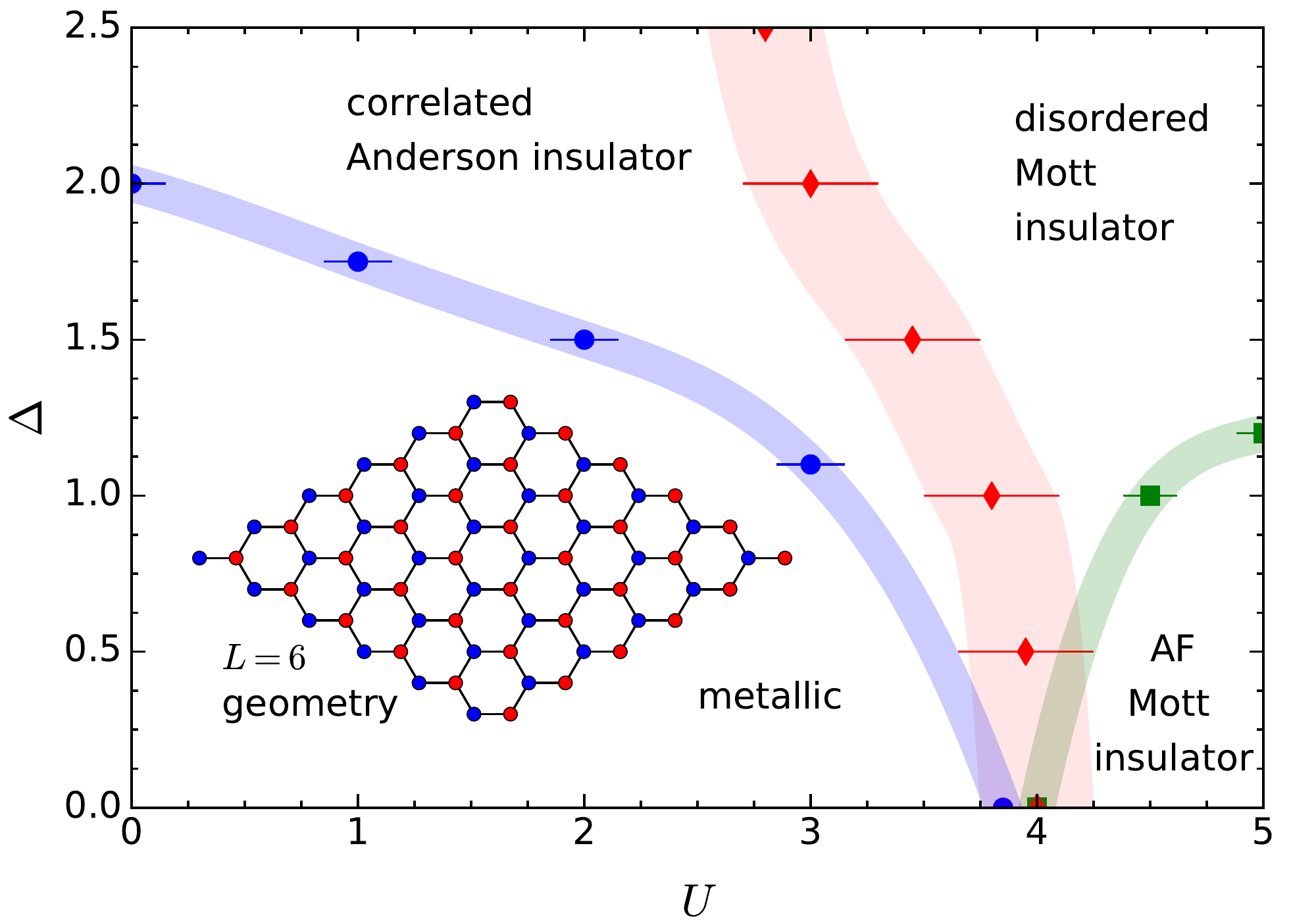}
\caption{Phase diagram of the disordered Hubbard model on the honeycomb
lattice at half-filling. $\Delta$ labels the disorder strength and $U$
represents the local Coulomb repulsion. 
The metallic phase boundary is determined by the temperature dependence
of the conductivity $\sigmadc$ and the region of long range AF order
by finite size scaling of the AF structure factor.
Although these transitions coincide in the clean limit,
for non-zero $\Delta$ an intermediate, magnetically disordered, insulator phase
intervenes.  This phase itself contains a crossover from Anderson-like
to Mott-like behavior based on the behavior of the compressibility
$\kappa$.
The inset shows the geometry of
the $L=6$ honeycomb lattice.  }
\label{Fig:PhaseDiagram}
\end{figure}

We focus on electronic, transport, as well as magnetic properties of the system, 
and highlight the phases that emerge from the competition of disorder and interaction.
The key results are summarized in the phase diagram Fig.~\ref{Fig:PhaseDiagram}: 
Whereas in the absence of disorder the metal-insulator and AF phase transitions 
coincide at a common critical coupling,\cite{Paiva2005,*Sorella2012,*Assaad2013,*Arya2015} 
the addition of disorder reduces the threshold coupling strength for insulating behavior, 
while increasing that for AF order, thereby opening up an intervening insulating phase 
with no long range magnetic order.

%Without disorder, the model has been the focus of many numerical
%studies.\cite{Paiva2005,*Sorella2012,*Assaad2013,*Arya2015} It is now
%established that the ground state of the model has a quantum critical
%point separating a paramagnetic semimetal and an antiferromagnetic Mott
%insulator. Similarly, transport in disordered graphene in the absence of 
%interactions has been extensively investigated
%\cite{Aleiner06,Fradkin86,Shon98,Suzuura02,Zheng02,Ando02,Khveschenko06,Morpurgo06,McCann06}.

%% A system with a quantum critical region separating a Fermi-liquid from
%% an antiferromagnetic Mott-insulator is the two dimensional
%% Fermi-Hubbard model on the honeycomb lattice, which offers the unique
%% opportunity to perform a detailed unbiased analysis of the effects of
%% quantum critical fluctuations on itinerant charge carriers. It has
%% recently attracted extensive studies due to its relevance for graphene,
%% silicene and germanene.

%% Moreover, in the semimetal region, the behavior of resistivity may be
%% divided into two kinds.  It is described by the steinhart$-$Hart
%% equation in low interaction strength, and interestingly, it behaves as
%% $\rho=ae^{b\//{T^2}}$ near but bellow the critical point.

% --------------------------------------------------------------------
\noindent
\underline{\it Model and numerical method} ---
We consider the bond disordered Hubbard Hamiltonian
\begin{align}
\hH = & -\sum_{\left\langle \bfi\bfj\right\rangle \sigma}
       t^{\phantom{\dagger}}_{\bfi\bfj} 
       \left( \hc^{\dagger}_{\bfi\sigma} \hc^{\phantom{\dagger}}_{\bfj\sigma} + 
       \hc^{\dagger}_{\bfj\sigma} \hc^{\phantom{\dagger}}_{\bfi\sigma} \right)
      -\mu\sum_{\bfi\sigma}\hn_{\bfi\sigma} \nonumber\\
    & + U\sum_\bfi \left( \hn_{\bfi\uparrow} - \frac 1 2\right) \left( \hn_{\bfi\downarrow} - \frac 1 2\right).
   \label{eq:model}
\end{align}
$\hc^\dag_{\bfi\sigma}$ ($\hc_{\bfi\sigma}$) are the spin-$\sigma$
electron creation (annihilation) operator at site $\bfi$. $U > 0$ is the
on-site Coulomb repulsion. $t_{\bfi\bfj}$ is the hopping integral
between two near-neighbor sites $\bfi$ and $\bfj$.
%%  (denoted by $\langle \bfi\bfj\rangle$ in the summation). 
The chemical potential $\mu$
determines the average density of the system.
$\hn_{\bfi\sigma}=\hc^{\dagger}_{\bfi\sigma}\hc^{\phantom{\dagger}}_{\bfi\sigma}$ is the number
operator.  Disorder is introduced through the hopping matrix elements
$t_{\bfi\bfj}$ chosen uniformly
$P(t_{\bfi\bfj})=1/\Delta$ for $t_{\bfi\bfj}\in
[t-\Delta/2,t+\Delta/2]$, and zero otherwise. Here $\Delta$ represents a
measure of disorder strength, and $t=1$ sets the scale of energy.  
In this work, we focus on the case
$\mu=0$ where the system is half-filled
and the Hamiltonian remains particle-hole
symmetric\cite{Denteneer1999} even in the presence of disorder.

The model is solved numerically using the finite-temperature DQMC
method.\cite{WhHbk} In this approach, the interacting Hamiltonian is
mapped onto free fermions coupled to space and imaginary-time dependent
Ising fields. The integration over all possible field configurations is
carried out by Monte Carlo sampling. This approach allows us to compute
static and dynamic (in imaginary time) observables at a given
temperature $T$.  Because of the particle-hole symmetry, the system is
sign-problem free and the simulation can be performed at large enough 
$\beta=1/T$ to converge to the ground state.  Data reported are obtained on
$2\times L^2$ ($L=3$, 6, 9, 12, and 15) honeycomb lattices with periodic
boundary conditions. The inset of Fig.~\ref{Fig:PhaseDiagram} shows the
$L=6$ geometry. In the presence of disorder, results are averaged over
20 disorder realizations; the error bars reflect both statistical and
disorder sampling fluctuations.

% --------------------------------------------------------------------

% --------------------------------------------------------------------
We use the temperature-dependent DC conductivity $\sigmadc(T)$ to characterize 
the metal-insulator transition.  According to the fluctuation-dissipation theorem, 
$\sigmadc(T)$ is related to the zero frequency limit of the current-current 
correlation function. While
real-frequency quantities can be obtained through analytic continuation
of imaginary-time QMC data, we implement an
approximation\cite{Trivedi1995} that has been extensively benchmarked in
previous work,\cite{Trivsig,Trivedi1995,Denteneer1999}
%% The DC conductivity is expressed as
\begin{equation}
 \sigmadc(T) =
   \frac{\beta^2}{\pi} \Lambda_{xx} ({\bf q}=0,\tau=\beta/2).
 \label{eq:condform}
\end{equation}
Here $\Lambda_{xx} ({\bf q},\tau) = \ob{ \hj_x ({\bf q},\tau) \, \hj_x
(-{\bf q}, 0)}$, where $\hj_x ({\bf q},\tau)$ is the Fourier transform of
the time-dependent current operator $\hj_x(\bfr,\tau)$ in the
$x$-direction.
%% \footnote{
%% The current operator is defined as\cite{Capone2001}
%% \begin{equation*}
  %% j_x (\bfr,\tau) =  e^{H\tau}\left( i\,t_{\bfr+\bfl,\bfr} \sum_{\sigma}\,l_x
  %% \left(c^{\dagger}_{\bfr+\bfl,\sigma} c_{\bfr\sigma} - \mathrm{h.c.} \right)
  %% \right)e^{-H\tau},
%% \end{equation*}
%% where $\bfl$ is the vector connecting near-neighbor sites.}
Eq.~(\ref{eq:condform}) provides a good approximation if the temperature
is lower than the energy scale at which 
there is significant structure in the density of states.\cite{Trivedi1995} 
%% It allows $\sigmadc$ to be extracted
%% directly from $\Lambda_{xx} ({\bf q},\tau)$. 
Checks of the applicability to the present problem will be discussed below.

In addition to transport properties, we also examine the charge
excitation gap and the antiferromagnetic (AF) structure
factor at wave vector $\bfQ=\Gamma$,
\begin{equation}
S_{AF}=\frac{1}{N_c}\sum_{\bf r} \left( \ob{\hat\bfS_{\bfr,A}}-\ob{\hat\bfS_{\bfr,B}} \right)^2.
\label{eq:Saf}
\end{equation}
Here $N_c$ is the number of unit cells.  ${\hat\bfS_{\bfr,A}}$ and
${\hat\bfS_{\bfr,B}}$ are total spin operators for sublattices $A$ and $B$
of the bipartite honeycomb lattice.

% --------------------------------------------------------------------
\begin{figure}
\includegraphics[scale=0.42]{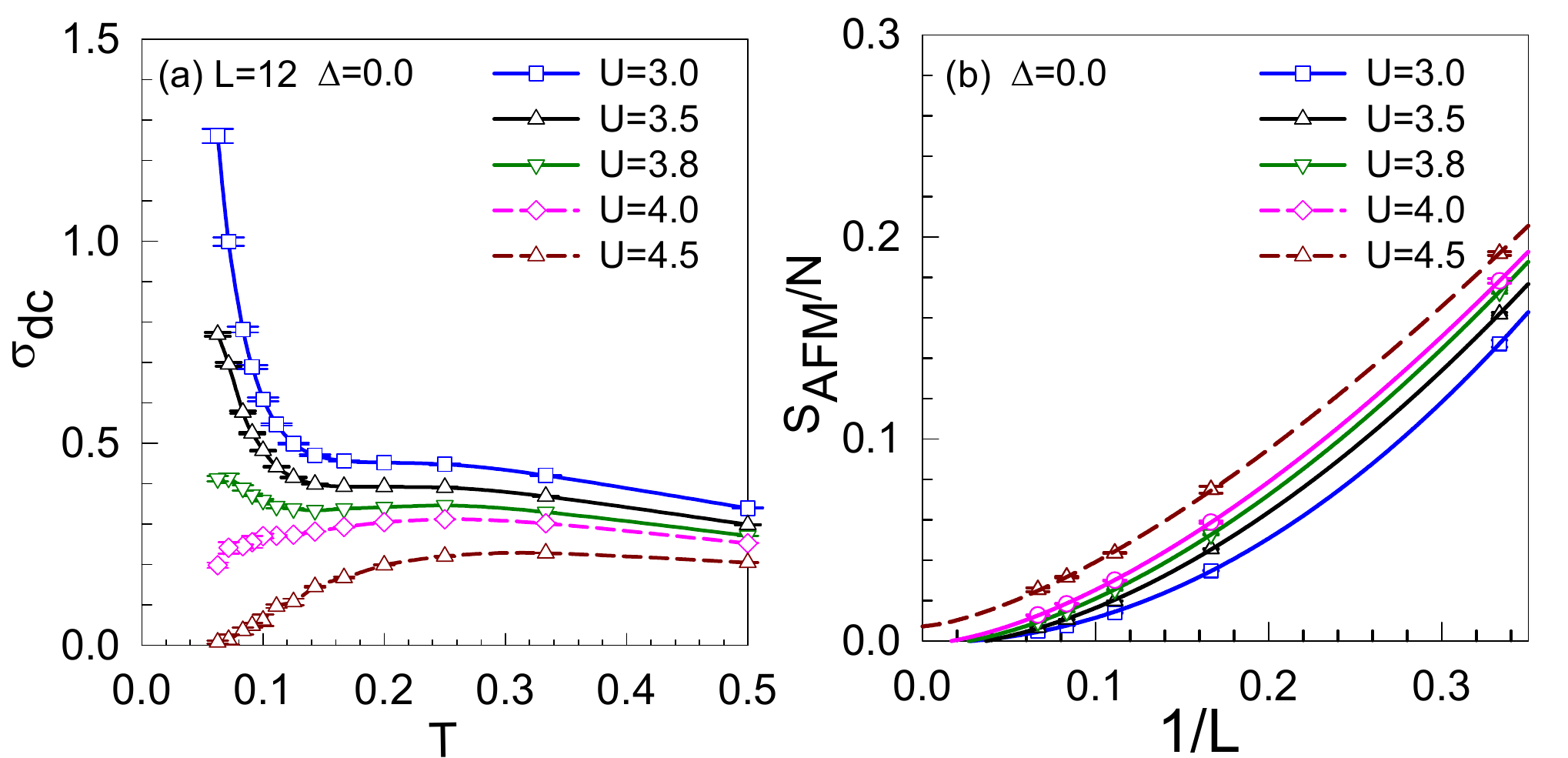}
\caption{(a) DC conductivity $\sigmadc$ versus temperature $T$ in the
clean limit $\Delta =0$ computed at various coupling strengths for the
$L=12$ honeycomb lattice. (b) Scaling behavior of the normalized AF spin
structure factor $S_{AF}/N_c$ at corresponding $U$ values. Solid and
dashed lines represent third-order polynomial fits to the data.
}
\label{Fig:sigmaDC.clean}
\end{figure}

% --------------------------------------------------------------------
\noindent
\underline{\it Results and discussion} ---
%% As a consistency check of Eq.~(\ref{eq:condform}) and (\ref{eq:Saf}), 
We first demonstrate results for the disorder-free system.
Fig.~\ref{Fig:sigmaDC.clean}(a) shows $\sigmadc(T)$  measured on the
$L=12$ lattice across several $U$ values. As shown by the figure, the
conductivity increases with decreasing temperature for $T \gtrsim 0.25$,
regardless of $U$. Upon further lowering $T$, the data indicate that
$d\sigmadc/dT < 0$ and $\sigmadc$ diverges as $T\rightarrow 0$  for
$U \lesssim 3.8$.  This low temperature behavior is an indication that
the system is metallic.\cite{Denteneer1999} For $U\gtrsim 4.0$, the
low-$T$ behavior of $\sigmadc$ points to an insulating state:
%% Specifically, below a characteristic temperature scale related to the
%% Mott gap, 
$d\sigmadc/dT > 0$ and the conductivity vanishes as
$T\rightarrow 0$.
%% The change of low-$T$ behavior suggests that there is a
%% metal-insulator transition at $3.8 \lesssim U\lesssim 4.0$.
%%
%% Regarding magnetic properties, 
Fig.~\ref{Fig:sigmaDC.clean}(b) shows
finite-size scaling of the normalized AF spin structure factor
$S_{AF}/N_c$. By extrapolating the data to the thermodynamic limit, it
appears that the onset of AF order is $3.8\lesssim U \lesssim 4.0$.
These findings suggest that there is a transition from paramagnetic
semimetal to an AF insulator at $3.8\lesssim U \lesssim 4.0$, a result
that is consistent with previous finding of a quantum
critical point $U_c\sim 3.85$ which separates the semimetallic and Mott
insulating phases.\cite{Otsuka2016}

% --------------------------------------------------------------------
\begin{figure}[b]
\includegraphics[scale=0.41]{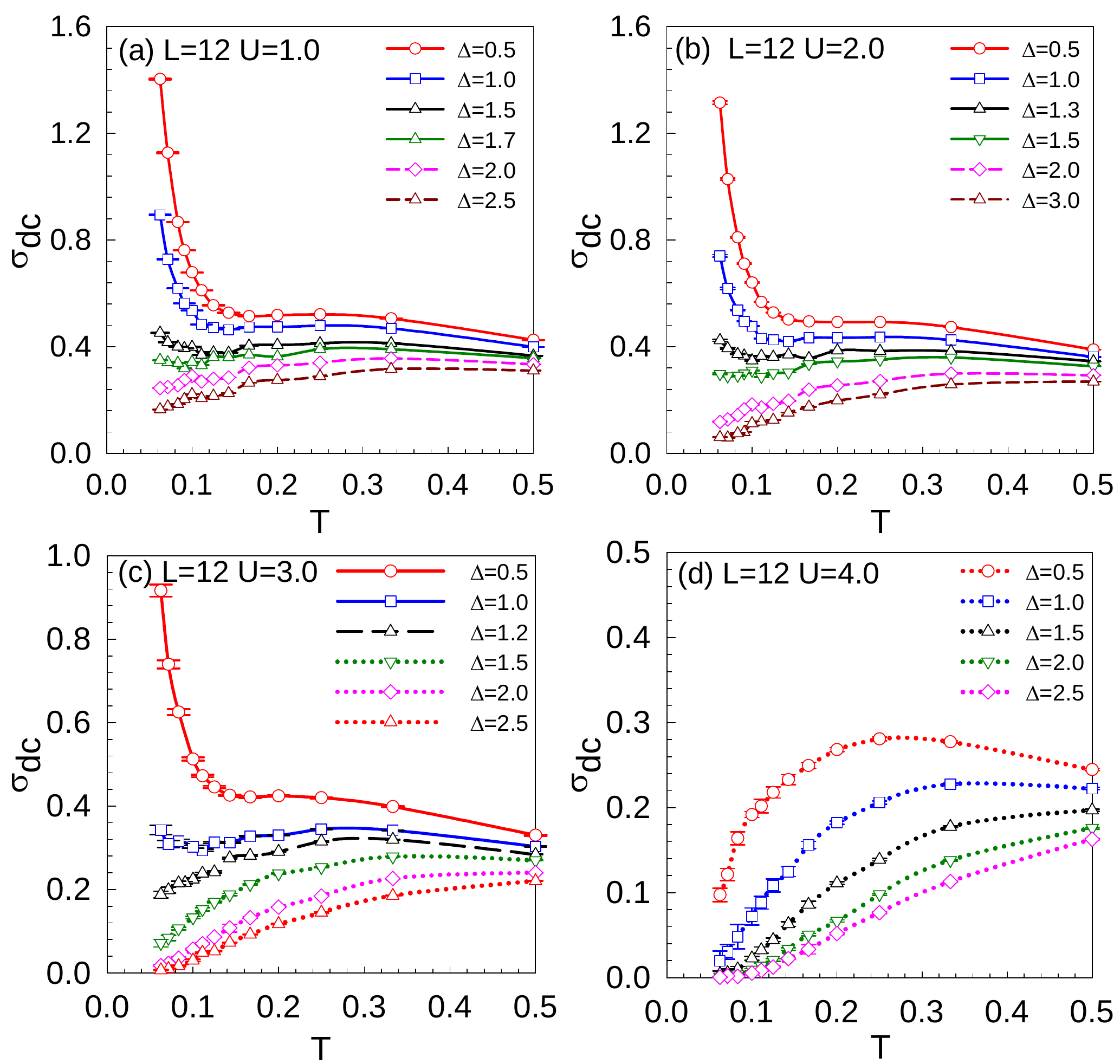}
\caption{Temperature dependence of the DC conductivity $\sigmadc$
measured on the $L=12$ lattice with disorder. Panels correspond to
different couplings: (a) $U=1.0$, (b) $U=2.0$, (c) $U=3.0$, and (d)
$U=4.0$. In each figure, lines are guides to the eyes.  Metallic and
insulating behaviors are indicated by solid and dashed lines
respectively.
}
\label{Fig:sigmaDC.disordered}
\end{figure}

%% In Fig. \ref{Fig:dcdis0} (b)  we show $\sigmadc(T)$ measured on
%% different lattice sizes $L=6$, 9, and 12 in the insulating regime
%% $U=4.0$. All data show a clear crossover from high-$T$ metallic to
%% low-$T$ insulating behavior. The crossover takes place at the same
%% temperature scale except for the $L=6$ lattice, possibly due to
%% finite-size effects. In the following, discussions will be based on
%% data obtained on the $L=12$ lattice.

% --------------------------------------------------------------------
\begin{figure}[b]
\includegraphics[scale=0.43]{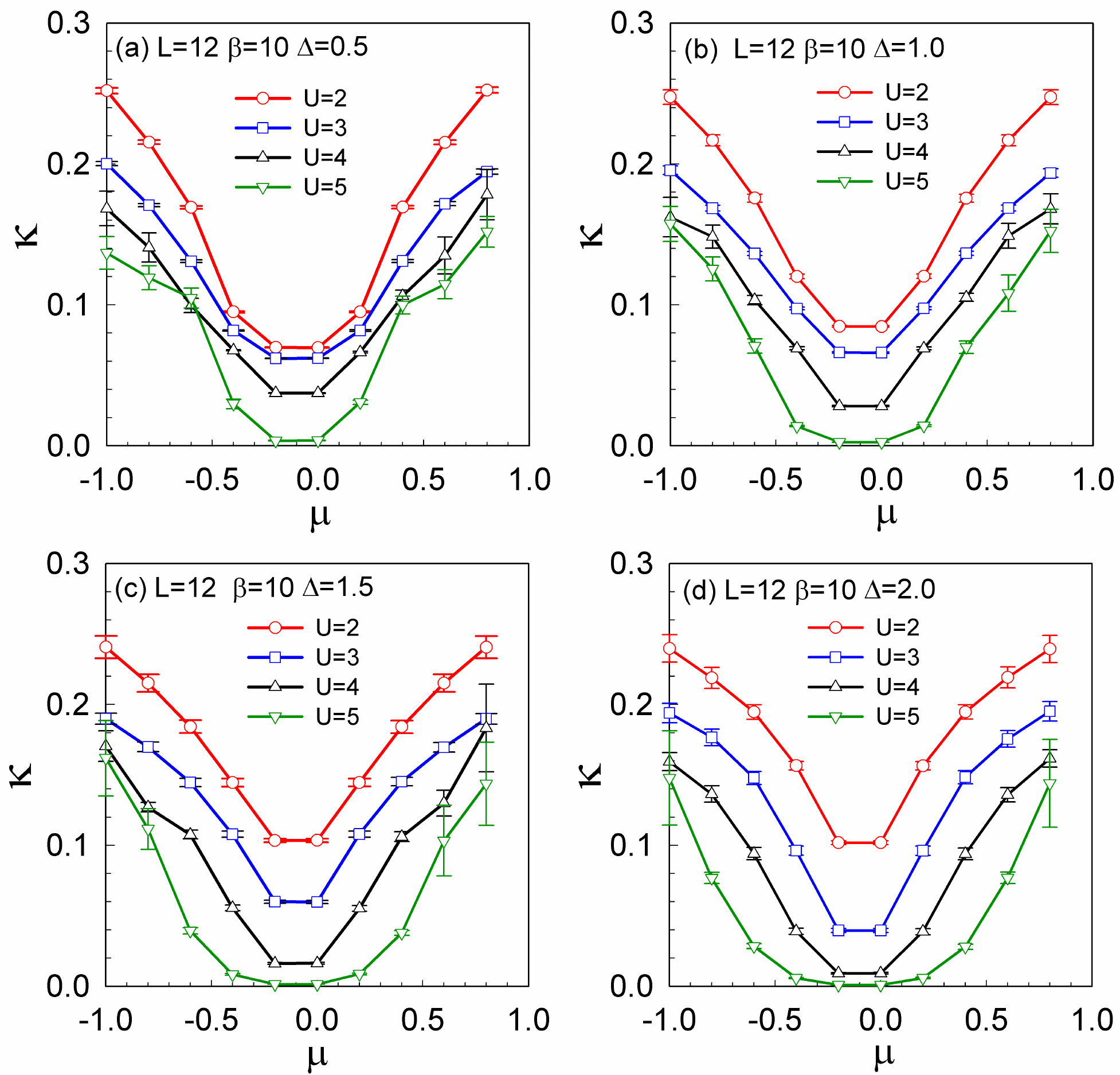}
\caption{Charge compressibility $\kappa$ versus chemical potential $\mu$ 
at four representative disorder strengths. A finite $\kappa$ means the
system is compressible, i.e., gapless; while $\kappa=0$ implies an
gapped incompressible state. The criterion $\kappa \lesssim 0.04$ 
is adopted to distinguish the gapped and gapless states.
}
\label{Fig:compressibility}
\end{figure}

% --------------------------------------------------------------------
\begin{figure}[t]
\includegraphics[scale=0.425]{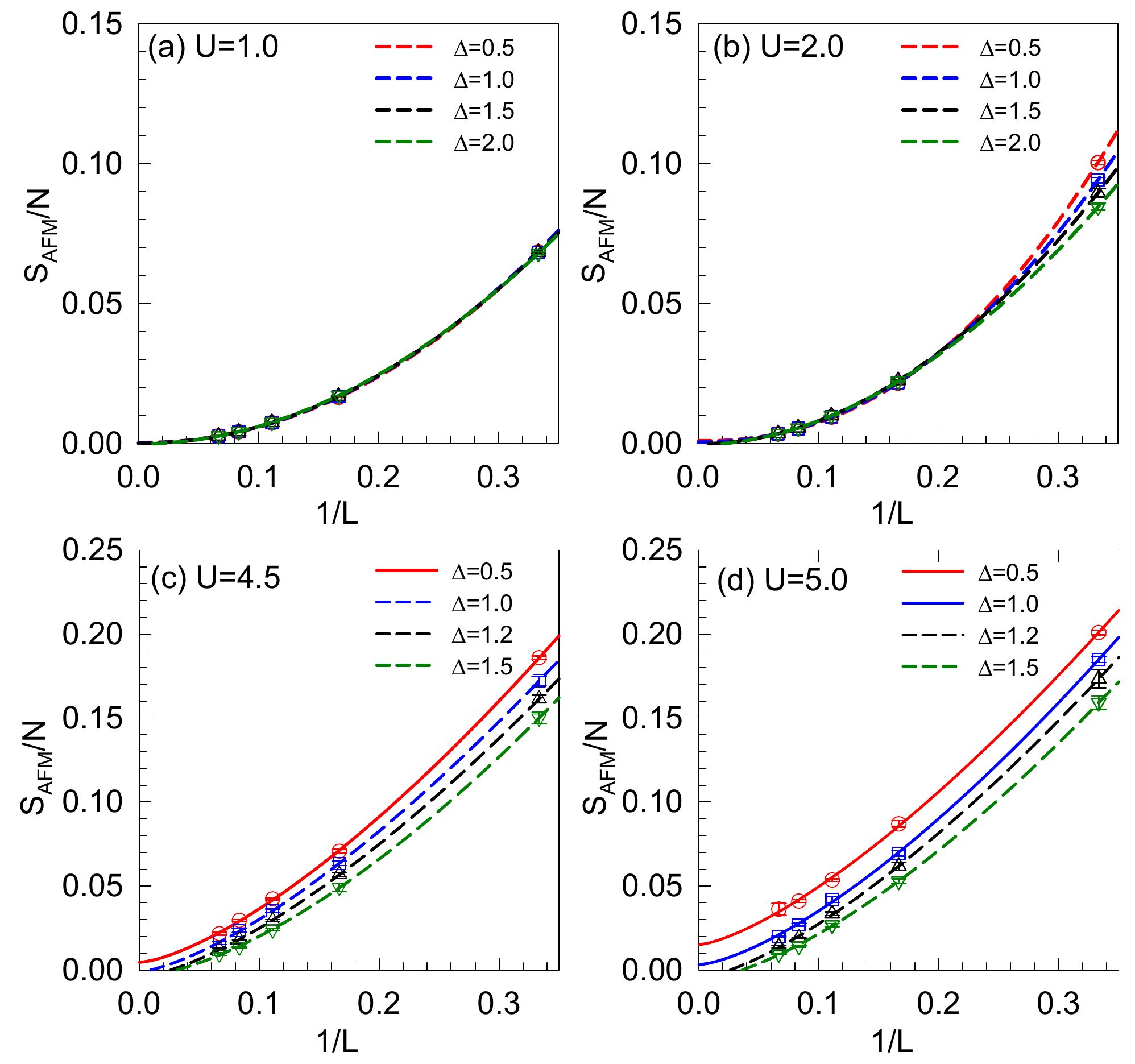}
\caption{Finite-size scaling studies of the AF spin structure factor.
In (c) and (d), antiferromagnetic spin structure factor, $S_{AF}$
is plotted as a function of $1/L$ for various value of disorder strength
at $U=4.5 t$ (c) $U=5.0 t$ (d); Lines are cubic polynomial (in $1/L$) fits to the data.
}
\label{Fig:Saf}
\end{figure}

Next we move to the disordered case. 
In disordered graphene and without interactions, electronic
transport has been extensively 
investigated.\cite{Aleiner06,Fradkin86,Shon98,Suzuura02,Zheng02,Ando02,Khveschenko06,Morpurgo06,McCann06}
In order to establish the phase diagram for interacting electrons, 
we first examine transport properties. 
Fig.~\ref{Fig:sigmaDC.disordered} shows $\sigmadc(T)$ computed in a 
range of disorder strengths
at four representative coupling strengths across the $\Delta=0$ quantum
critical point $U_c$.

Fig.~\ref{Fig:sigmaDC.disordered}(a) $\sim$ (c) examine the semi-metallic 
region $U\lesssim U_c$. The low
temperature behavior of $\sigmadc$ indicates that there is a change from
metallic to insulating behavior with increasing disorder $\Delta$: For
$U=1$, $T\lesssim 0.14$, and $\Delta=0.5$, the conductivity is metallic.
$\sigmadc(T)$ grows with decreasing temperature: $d\sigmadc/dT < 0$.  At
$\Delta=2.5$, on the other hand, $\sigmadc(T)$ decreases as the
temperature is lowered, and approaches zero as $T\rightarrow
0$, suggesting insulating behavior. The crossover from a metallic to an
insulating state takes place at $\Delta_c\sim 1.7$ for $U=1$.  By
raising the interaction strength, the crossover sets in at a reduced
disorder strength: $\Delta_c\sim 1.5$ and {\color{blue}1.0} for $U=2$ and $3$
respectively. The critical disorder strength reduces to $\Delta=0$ at
roughly $U\sim 3.9$ where the system enters the correlation-induced
Mott-Slater insulator regime. The conductivity data exhibit an
insulating response $d\sigmadc/dT > 0$ and vanish as $T\rightarrow 0$
for any $\Delta$. See Fig.~\ref{Fig:sigmaDC.disordered}(d).

The ``metallic'' region of the phase diagram Fig.~\ref{Fig:PhaseDiagram}
summarizes these transport results.  As previously found for the
quarter-filled square lattice Hubbard model \cite{Denteneer1999} with
bond disorder, our DQMC calculations suggest that the onsite Hubbard
repulsion can introduce metallic behavior in the 2D honeycomb lattice
even at the Dirac point where the density of states $N(E_F)=0$ for $U=0$.

Another electronic property of interest is the single-particle gap.
Without disorder, the half-filled Hubbard model on the honeycomb lattice
exhibits a charge (Mott) excitation gap for $U >
U_c$.\cite{Assaad2013,Otsuka2016} The non-interacting Anderson
insulator, on the other hand, is gapless at the Fermi level (in the
thermodynamic limit).\cite{Anderson1978,Thouless2010} Although the
gap is not an order parameter associated with symmetry
breaking, it nevertheless can be used to establish the existence of the
Mott insulator.  
%The vanishing of the gap when both disorder and
%interaction are present is quite subtle,\cite{Shinaoka2009,Byczuk2010}.
%
In general the single-particle gap can be extracted from the density
of states $N(\omega)$. Here we extract information of the gap by examining 
the behavior of charge compressibility 
$\kappa = -d\ob{\hn(\mu)}/d\mu$ at the Fermi level $\mu=0$, where $\ob{\hn(\mu)}$ is the
average density at the chemical potential $\mu$.
Using this formula, the compressibility
can be deduced from local densities which are easy to compute within DQMC.
A finite $\kappa$ indicates that the system is compressible, i.e., gapless.

In Fig.~\ref{Fig:compressibility}, the compressibility $\kappa$ is plotted 
as a function of $\mu$ for various disorder strength $\Delta$ and local
repulsion $U$. Each data point is obtained by averaging results from 20 disorder 
realizations on the $L=12$ lattice. 
Tuning the chemical potential away from $\mu=0$ breaks particle-hole symmetry 
and leads to a sign problem. However, the problem is less severe in the presence 
of disorder\cite{Denteneer1999,Paiva2015}, and we are still able to extract accurate 
local denisty results.  
In the weak disorder region $\Delta \ll U$, Fig.~\ref{Fig:compressibility} indicates
that $\kappa$ vanishes near $\mu=0$ for $4.0 \lesssim U \lesssim 5.0$, i.e., the 
system becomes incompressible and acquires an energy gap for charge excitations.  
For strong disorder $\Delta\gtrsim U$, it is clear that the compressibility tends to
become zero at a weaker coupling strength $3.0 \lesssim U \lesssim 4.0$. 
We are not able to pin-point the exact location where the gap opens at each disorder 
strength due to the sparse data. Nonetheless, an estimated cross-over separating the 
gapless and gapped regions, using the criterion $\kappa \lesssim 0.04$, is
presented in Fig.~\ref{Fig:PhaseDiagram}.

We now consider magnetic properties and the effect of bond disorder on long-range AF, 
generalizing the discussion of Fig.~\ref{Fig:sigmaDC.clean}(b). Fig.~\ref{Fig:Saf} 
summarizes finite-size scaling studies of the  AF structure factor on lattices up to 
$L=15$ (450 sites). For $\Delta=0$, it is known that the ground state of the system
is an antiferromagnet for $U > U_c\sim 3.85$.\cite{Otsuka2016} For $U=1,2$, where there 
is no AF order even in the clean limit, bond disorder has almost no effect on $S_{AF}$ 
for sufficiently large lattices (Figs.~\ref{Fig:Saf}(a) and (b)). On the other hand, 
above the clean limit $U_c$, $\Delta > 0$ suppresses $S_{AF}$ and increases the interaction
strength needed for long range AF order to appear. The mechanism for the suppression 
of AF is the tendency towards singlet formation on pairs of sites with large 
$t_{\bfi\bfj}$\cite{Enjalran2001}.
Based on the extrapolated behavior of $S_{AF}/N_c$ in the thermodynamic limit, a 
magnetic phase boundary for the paramagnetic-antiferromagnetic transition can be 
established for nonzero $\Delta$, and is shown in Fig.~\ref{Fig:PhaseDiagram}.

\noindent
\underline{\it Summary} ---
We have studied electronic and magnetic properties of the disordered Hubbard model 
on the honeycomb lattice using DQMC simulations. In the absence of disorder, we 
verified this geometry has a quantum critical point at $3.8 \lesssim U\lesssim 4.0$ 
separating the semimetallic and Mott insulating phases, a result that is consistent
with previous (higher resolution) findings.\cite{Sorella2012,Assaad2013} 
In the $U=0$ limit, the semimetallic phase is driven into an gapless Anderson 
insulating state. By switching on the local Coulomb repulsion $U$, the critical 
disorder strength for the metal-insulator transition decreases, indicating that the 
presence of both disorder and interactions becomes more effective in localizing electrons.
At $U \gtrsim 4.0$, electrons are localized by strong Coulomb correlations in the 
absence of disorder: the AF transition and metal-insulator transitions coincide in
the clean limit. Our key finding is that adding random bond disorder reduces the 
threshold $U$ required for insulating behavior, but increases the $U$ required for AF.
Thus the magnetic and metal-insulator transitions no longer coincide, and a disordered 
insulating phase intervenes. Furthermore, within this disordered insulator, there is 
a crossover from an Anderson-like region where the compressibility $\kappa \neq 0$
to a Mott-like region where $\kappa=0$.

Already, certain unique features of the interplay of disorder and interactions in models 
with a Dirac dispersion have been noted, including the possibility that disorder might 
enhance superconductivity for attractive interactions\cite{Potirniche14}. Our work expands 
this understanding to repulsive interactions, where similar anomalous effects such as an 
enhancement of N\'eel temperature by randomness are known\cite{Ulmke95} for conventional geometries.
Moreover, the reduced critical coupling strength for the metal-insulating transition 
in the presence of disorder might be relevant for practical applications of honeycomb 
structural materials such as a low power Mott transistor.

\noindent
\underline{\it Acknowledgement} ---
T. M. thanks CAEP for partial financial support. T. M. and L. F. Z were
supported by NSFCs (Grant.~Nos. 11374034 and 11334012), and the
Fundamental Research Funds for the Center Universities, grant
No.~2014KJJCB26. We acknowledge computational support from the
Beijing Computational Science Research Center (CSRC), the support of
HSCC of Beijing Normal University, and phase 2 of the Special Program
for Applied Research on Super Computation of the NSFC-Guangdong Joint
Fund.  R. T. S. was funded by the Department of Energy (DOE)
under Grant No. DE-NA0001842-0.  C.-C. C. acknowledges DOE-LLNL
support under Contract DE-AC52-07NA27344, 15-ERD-013.

\bibliography{reference}

\end{document}